\gdef\journal#1, #2, #3, 1#4#5#6{		
    {\sl #1~}{\bf #2}, #3 (1#4#5#6)}		

\def\prd{\journal Phys. Rev. D, }

\def\prl{\journal Phys. Rev. Lett., }

\def\cmp{\journal Comm. Math. Phys., }

\def\np{\journal Nucl. Phys., }

\def\pl{\journal Phys. Lett., }

\def\annp{\journal Ann. Phys. (N.Y.), }

\def\ijmpA{\journal Int. Jour. of Mod. Phys. A, }

\documentstyle[12pt]{article}
\begin{document}
\renewcommand{\theequation}{\thesection .\arabic{equation}}
\newcommand{\beq}{\begin{equation}}
\newcommand{\eeq}{\end{equation}}
\newcommand{\beqn}{\begin{eqnarray}}
\newcommand{\eeqn}{\end{eqnarray}}
\newcommand{\slp}{\raise.15ex\hbox{$/$}\kern-.57em\hbox{$\partial$}}
\newcommand{\lnA}{\raise.15ex\hbox{$/$}\kern-.57em\hbox{$A$}}
\newcommand{\lnB}{\raise.15ex\hbox{$/$}\kern-.57em\hbox{$B$}}
\input{jnl.modif}

\begin{titlepage}

\begin{center}
{\large{\bf Zamolodchikov's c-function for the Chiral Gross-Neveu Model}}
\end{center}
\vspace{1.5cm}
\begin{center}
D.C.Cabra$^{a}$
\end{center}
\begin{center}
Depto. de F\'{\i}sica. Universidad Nacional de la Plata
\end{center}
\begin{center}
La Plata. Argentina.
\end{center}
\vspace{1.5cm}

ABSTRACT:{\small  We construct the Zamolodchikov's c-function for the Chiral
Gross-Neveu Model up to two loops. We show
that the c-function interpolates between
the two known critical points of the theory, it is stationary at them and
it decreases with the running coupling constant.
In particular one can infer the non-existence
of additional critical points in the region under investigation.}
\vspace{5cm}

\noindent --------------------------------

\noindent $^a$ {\footnotesize Consejo Nacional de Investigaciones
Cient\'\i ficas y T\'ecnicas, Argentina.\\
Postal Address: Depto. de F\'\i sica. Universidad Nacional de
La Plata.\\ CC 67, 1900 La Plata, Argentina.
Email Address: cabra@fisilp.edu.ar}

\pagebreak

\section{Introduction and Results}

In the last years, two-dimensional conformal field theories (CFT) have been
the subject of a lot of research, after the pioneering works of Polyakov
and Belavin, Polyakov and Zamolodchikov\cite{Pol}.
The interest come because statistical
mechanical models at the critical point are conformally invariant and also
because CFT correspond to ground states of string theories\cite{Wi1,Cdy2}.

The understanding of 2D CFT has been enhanced after the formulation
of the Zamolodchikov's c-theorem which provides
a powerful tool in the analysis of two-dimensional field theories
away from criticallity\cite{16}.
This theorem states that there exists a scalar function $c(g)$ of the couplings
$g$ of the theory, which behaves monotonically under a Renormalization Group
flow (decreasing into the infrared) and whose value at a fixed point (where
the theory is conformally invariant),
is the central charge of the corresponding Virasoro Algebra.

The interest in the theorem relies in the fact that it could give
non perturbative information
of a non-critical theory from certain well known properties of the theory at
criticallity (where it could be in principle exactly solved, see
Ref.\cite{Cdy})
Moreover, after a reformulation
by Friedan\cite{Frie} in terms of spectral functions, it is possible to
study generalizations of the Zamolodchikov's c-theorem in four dimensions.
Indeed, there have been many attempts in this direction\cite{Frie,attemps}
and several
candidates have been proposed to play the role of the $c(g)$-function in
four dimensions.

It is then clear that a depeer understanding of the
c-theorem should
help both in the study of two-dimensional statistical systems and also
in the construction of generalizations relevant to the study of four
dimensional CFT.

An explicit calculation of the Zamolodchikov's c-function have been already
presented
for the (exactly solvable) Schwinger Model\cite{chinos}, but the construction
of the c-function in a field theoretical model with  two non-trivial fixed
points is lacking. It is the purpose  of the present work to fill this gap by
constructing the c-function up to two loops for the Chiral
Gross-Neveu (CGN) model\cite{GN}.

The interest in the CGN model originates in the fact that it posseses (given
a suitable regularization scheme) two non-trivial fixed points at which
the theory becomes conformally invariant.
These critical points were shown to exist in Refs.\cite{DF}-\cite{DdV}.
Moreover, in Ref.\cite{2} (using path-integral techniques) the Virasoro
Central Charge (VCC) {\cal at} the fixed points was evaluated.

Our purpose is then to test Zamolodchikov's c-theorem in an explicit
non-trivial
example, showing that all properties predicted in its thesis are satisfied.
The main results in the present work are the following: we show that
Zamolodchikov's c-function for the CGN model, up to two loops,
interpolates between these points and is stationary at them. Moreover,
it is shown that $c(g)$ has
no additional stationary points between those above mentioned, thus confirming
that the model becomes conformally invariant only at the critical points
found in Ref.\cite{2}.
The resulting $c$-function is monotonically decreasing
with the running coupling constant
$g^2_N$ (for a given regularization scheme) and then we can guess the sign
of the $\beta$-function that must be positive in this domain.

The plan of the paper is as follows:
a brief description of the model is given in Section II.
In Section III we describe the evaluation of the one and two loop contributions
to the Zamolodchikov's c-function. At the end of this section we discuss
our results and its possible extensions to the study of Coset Models.
Details on the calculations are left to an Appendix.

\section{The Chiral Gross-Neveu Model}

\indent We start from the Lagrangian (in two-dimensional Euclidean space):
\beq
{\cal L}=\overline{\Psi}^i i\slp \Psi^i
-\frac{1}{2}g^2_sj^{\mu}j_{\mu}-\frac{1}{2}g^2_Nj^{\mu a}j_{\mu}^a ,\label{2.1}
\eeq
where $\Psi^i ~(i=1,...,N)$ are Dirac fermions and $j^{\mu}$ and $j^{\mu
a}$ are U(1) and SU(N) currents:
\beqn
j^{\mu}&=&\overline{\Psi}^i\gamma^{\mu}\Psi^i \nonumber \\
j^{\mu a}&=&\overline{\Psi}^i\gamma^{\mu}t^a_{ij}\Psi^j   \label{2.2}
\eeqn

\noindent with $t^a~(a=1,...,N^2-1)$ the SU(N) generators.

Our conventions are:

\beqn
[t^a,t^b]&=&2if^{abc}t^c, \nonumber   \\
tr(t^at^b)&=&2\delta^{ab},  \nonumber   \\
t^at^a&=&c_dI,    \nonumber   \\
f^{abc}f^{a'bc}&=&c_A\delta^{aa'}.  \label{2.3}
\eeqn

Following \cite{1}, we introduce auxiliary fields $A_{\mu}^a$ and $B_{\mu}$
in order to eliminate quartic interactions.
After this is done, the partition function reads:

\beq
{\cal Z}=\int D\overline{\Psi} D\Psi DA^a_{\mu}DB_{\mu}
exp\left(-\int d^2x{\cal L}'[\overline{\Psi}, \Psi,A^a_{\mu}, B_{\mu}]\right)
\label{2.4}
\eeq
whith ${\cal L}'$ :

\beq
{\cal L}'=\overline{\Psi}^i(i\delta_{ij}\slp -g_s\lnB \delta_{ij}
-g_N\lnA^a t^a_{ij})\Psi+\frac{1}{2}B_{\mu}B^{\mu}
+\frac{1}{2}A_{\mu}^aA^{\mu a} \label{2.5}
\eeq

As it is well known, fermions can be completely decoupled from the auxiliary
fields\cite{1}: In order to decouple the U(1) field $B_{\mu}$, we perform
the following change of variables,

\beqn
\Psi&=&e^{(i\eta+i\gamma_5\phi)} \chi  \nonumber \\
\overline{\Psi}&=&\overline\chi e^{(-i\eta+i\gamma_5\phi)} \nonumber \\
B_{\mu}&=&\frac{1}{g_s}(\epsilon_{\mu\nu}\partial^{\nu}\phi-\partial_{\mu}\eta),
\label{2.6}
\eeqn
with the resulting partition function:

\beqn
\lefteqn{{\cal Z}=\int D\overline\chi D\chi DA^a_{\mu}D\phi D\eta J_B^{U(1)}
J_F^{U(1)}}
\nonumber \\
& & exp\left(-\int d^2x (\overline\chi^i(i\delta_{ij}\slp
-g_N\lnA^at^a_{ij})\chi+\frac{1}{2}A_{\mu}^aA^{\mu a})\right) \nonumber \\
& & exp\left(-\int d^2x\frac{1}{2g_s^2}
(\partial_{\mu}\eta\partial^{\mu}\eta+\partial_{\mu}\phi\partial^{\mu}\phi)
\right).
\label{2.7}
\eeqn

The jacobians $J_B^{U(1)}$ and $J_F^{U(1)}$, associated with the change of
variables (\ref{2.6}), are given by \cite{2}:

\beqn
J_B^{U(1)}&=&det(-\nabla)  \nonumber \\
J_F^{U(1)}&=&exp\left(-\frac{N}{2\pi}\int d^2x \partial_{\mu}\phi\partial^{\mu}
\phi\right)
exp\left(-\frac{b}{2\pi}\int d^2x B_{\mu} B^{\mu}\right).
\label{2.8}
\eeqn

Here $b$ is an undetermined parameter, related to regularization
ambiguities \cite{4} (it is on gauge invariance grounds that one usually
imposes $b=0$).

Concerning the SU(N) auxiliary field $A_{\mu}$, the decoupling change
of variables is:

\beqn
\chi_L&=&g^{-1}\chi_L^{(0)}~~~~~\chi_R=h^{-1}\chi_R^{(0)} \nonumber \\
{}~~~~~~~~~~~~~~~~ \nonumber \\
 \overline\chi_L&=&\overline\chi_L^{(0)}g~~~~~~~~\overline\chi_R=
 \overline\chi_R^{(0)}h, \nonumber \\
{}~~~~~~~~~~~~~~~~ \nonumber \\
&~&A_{+}= -(\frac{i}{g_N})g^{-1}\partial_{+} g , \nonumber \\
&~&A_{-}= -(\frac{i}{g_N})h^{-1}\partial_{-} h, \label{2.9}
\eeqn
where $\chi_L(\chi_R)$ are the left-handed (right-handed) fermionic components,
and  $g$ and $h$ are SU(N)-valued fields. After this change we get:

\beq
{\cal Z}= det(-\nabla) (det(i\slp))^N~{\cal Z}_{I}{\cal Z}_{II},
\label{2.10}
\eeq
where ${\cal Z}_{I}$ corresponds to a free bosonic theory
\beq
{\cal Z}_{I}=\int D\phi D\eta
exp\left(\frac{([b+\pi]-N)}{g_s^2}\partial_{\mu}\phi\partial^{\mu}\phi-
\frac{(b+\pi)}{g_s^2} \partial_{\mu}\eta\partial^{\mu}\eta\right),
\label{2.11a}
\eeq
and ${\cal Z}_{II}$ defines the partition function:
\beq
{\cal Z}_{II}=\int Dg Dh J(g) J(h) J_F^{SU(N)}
exp\left(\frac{1}{4g_N^2} \int d^2x tr(g^{-1}\partial _{+}g
h^{-1}\partial_{-}h)\right). \label{2.11b}
\eeq
In eq.(\ref{2.11b}), as it is well known \cite{3}, the jacobians associated
to the transformation (\ref{2.9}) are given by:

\beq
J_F^{SU(N)}=\frac{det(i\slp -g_N\lnA)}{det(i\slp)}=
exp\left(W[gh^{-1}]+\alpha\int d^2x
tr(g^{-1}\partial_{+}gh^{-1}\partial_{-}h)\right),
\label{2.12}
\eeq
and

\beq
J(g)= detD[A_+]^{Adj}= [J_F^{SU(N)}(A_{+})]^{2c_A} (det(i\slp))^{Adj} ,
\label{2.13}
\eeq
where the superscript $"Adj"$ means that determinants correspond to operators
in the adjoint representation.

In (\ref{2.12}), $W[g]$ is the Wess-Zumino-Witten (WZW) action\cite{3,Wi}:

\beq
W[g]=-\frac{1}{8\pi} \int
d^2xtr(\partial_{\mu}g\partial^{\mu}g^{-1})-\frac{i}{12\pi}
\int d^3y\epsilon^{ijk}
tr(g^{-1}\partial_igg^{-1}\partial_jgg^{-1}\partial_kg).
\label{2.14}
\eeq

Again, regularization ambiguities in $J_F^{SU(N)}$\cite{3}, \cite{6}, \cite{9}
are taken into
account by an arbitrary parameter $\alpha$.

Expression (\ref{2.10}) (already obtained in \cite{2})
gives a factorized form
for the partition function, which will be the starting point of our analysis
of the Zamolodchikov's c-function for the CGN Model.

Let us first review the conformal properties of the model:
the two determinants in (\ref{2.10}) arising from ghost fields and free
fermions, correspond to conformally invariant field theories with Virasoro
central charges given by\cite{dino}:
\beqn
c_{ghosts}&=&-2, \nonumber \\
c_{free fermions}&=&N. \label{2.15'}
\eeqn
Concerning ${\cal Z}_{I}$, it corresponds to the partition function of two
free boson fields, $\phi$ and $\eta$, thus contributing with a central
charge:

\beq
c_{bosons}=2.
\label{2.15''}
\eeq

Finally, ${\cal Z}_{II}$  corresponds to interacting SU(N)$_k$-fields
$g$ and $h$ and in general it does not correspond to a conformally invariant
model.

In fact, in Ref.\cite{2} it was shown  that only for certain relations
between the regularization
ambiguity parameter $\alpha$ and the coupling constant $g_N^2$, the theory
is conformally invariant.
The idea in Ref.\cite{2} is to adjust $\alpha$ and $g_N^2$ so that the
CGN Model partition function becomes a product of free (boson
and fermion) fields partition functions and WZW models (at the critical
point) partition functions so that each factor corresponds to a conformally
invariant model. To see this, we write ${\cal Z}_{II}$ in (\ref{2.11b})
in the following form:

\beq
{\cal Z}_{II}= \int Dg Dh exp(-kW[g]-kW[h^{-1}]+\frac{k}{4\pi}\tilde{g}\int
d^2x
tr(g^{-1}\partial_{+}gh^{-1}\partial_{-}h))
\label{2.16}
\eeq
where we have made use of the identity\cite{3}:

\beq
W[gh^{-1}]= W[g]+W[h^{-1}]-\frac{1}{4\pi}\int d^2x
tr(g^{-1}\partial_{+}gh^{-1}\partial_{-}h)
\label{2.17}
\eeq
and defined:

\beqn
k&=&-(2N+1) \nonumber \\
\tilde{g}&\equiv&\frac{\pi}{kg_N^2}-(\frac{4\pi\alpha}{k}+
1).
\label{2.18}
\eeqn
In what follows, $\tilde{g}$ will play the role of the effective coupling
constant of the model.

One now easily sees that
there are two fixed points for which the theory becomes conformally invariant.
Indeed, for $g_N^2=g_N^{2*}$ such that
\beq
g_N^{2 *}=\frac{1}{4k\alpha}
\label{2.19}
\eeq
the effective coupling constant $\tilde{g}$ becomes $\tilde{g}^{*}=-1$,
and at this point, the partition function can be written as:

\beq
{\cal Z_{II}}=\left[\int Dg\right] \int DU exp(-kW[U^{-1}]),
\label{2.19a}
\eeq
where $U=g.h^{-1}$.

Appart from the $\int Dg$ factor,
the partition function ${\cal Z_{II}}$ becomes that of a WZW model at the
critical point and hence ${\cal Z_{II}}$ (and then the complete partition
function ${\cal Z}$ defined by eq.(\ref{2.10})) corresponds to a conformally
invariant theory.

The other critical point is attained when $g_N^2=g_N^{2**}$ such that:

\beq
g_N^{2 **}=\frac{1}{4k[\alpha+1/4\pi]},
\label{2.20}
\eeq
for which $\tilde{g}^{**}=0$.
In this case, we have two decoupled SU(N)$_k$-WZW models at the critical
point. Again ${\cal Z}$ defines at this point a conformally invariant theory.

The corresponding values of the Virasoro central charge
for the WZW-sector (\ref{2.16}) at the
critical points (\ref{2.19}) and (\ref{2.20}), are given respectively
by:

\beqn
c^{*}&=&\frac{k~dimG}{k+c_A},  \label{2.21a} \\
c^{**}&=&2~\frac{k~dimG}{k+c_A}.
\label{2.21b}
\eeqn

Adding to this values the contributions comming from free fields
(\ref{2.15'}) and
(\ref{2.15''}), the total central charge $c_{total}$ associated to ${\cal
Z}$ at the critical points is:

\beq
c_{total}=N+\left\{
\begin{array}{lc}
\mbox{$\frac{kdimG}{k+c_A}$}&if~ g_N^{2}=g_N^{2 *} \\
{}~~~~&~~~~~\\
{}~~~~&~~~~~\\
\mbox{$2\frac{kdimG}{k+c_A}$}&if~g_N^{2}=g_N^{2 **}
\end{array}\right. \label{2.22}
\eeq

This analysis has not exhausted the possibility of having other critical
points at which the CGN model becomes conformally invariant. Precisely in
the next section we shall analize the Zamolodchikov's $c(\tilde{g})$-function
where
the effective coupling constant $\tilde{g}$ is  defined in eq.(\ref{2.18}).
Indeed for those values
of $\tilde{g}=\tilde{g}_i$ at which
$\frac{dc}{dg}|_{\tilde{g}_i}\propto \beta(\tilde{g}_i)=0$ the
model becomes conformally invariant. As we shall show such a condition is
not fullfilled by any value of $\tilde{g}$ in the interval
$(\tilde{g}^{*},\tilde{g}^{**})$.

\section{Zamolodchikov's c-function for the Chiral Gross-Neveu model}
\setcounter{equation}{0}

In this Section we are going to construct the Zamolodchikov's c-function
\cite{16} associated with the CGN Model.
In \cite{16}, Zamolodchikov devised a function $c(\tilde{g})$ depending on
the coupling constants (which we call $\tilde{g}$) of two-dimensional
theories showing that this function is:
\begin{itemize}
\item{a) non-negative and
non-increasing along a renormalization group trayectory,}
\item{b) its stationary points corresponds to critical fixed
points of the theory}
and
\item{c) at these fixed points, $c(\tilde{g})$ equals the Virasoro
central charge of the theory}.
\end{itemize}

In general, Zamolodchikov's c-function takes the form\cite{16}:

\beqn
\lefteqn{c(\tilde{g})=[2z^4<T(x)T(0)>} \nonumber \\
& &+4z^2x^2<T(x)\Theta(0)>-6x^4<\Theta(x)\Theta(0)>]
|_{x^2=R^2}, \label{4.9}
\eeqn
where $z=x_0+ix_1,~\overline{z}=x_0-ix_1$ and, (see eq.(2.20):
\beq
\tilde{g}=-\frac{4\pi}{k}\left(
k(\frac{1}{4\pi}+\alpha)-\frac{1}{4g_N^2}\right),
\label{4.6b}
\eeq
$T=T_{zz}$ and $\Theta=4T_{z\overline{z}}$ the two independent
components of the energy momentum tensor in these coordinates.
($R$ is a normalization point).
Properties a), b) and c) can then be proved using euclidean invariance,
unitarity and renormalizability\cite{16}.

We have shown in the precedent section that the CGN Model has two
critical points (corresponding to $g_N^2=g_N^{*2}$ ($\tilde{g}=-1$) and
$g_N^2=g_N^{**2}$ ($\tilde{g}=0$) ,
see eqs.(\ref{2.19}),(\ref{2.20})) at which the model is conformally invariant.
We shall construct in this section the $c(\tilde{g})$ function interpolating
between
these two points using Zamolodchikov's construction. The knowledge of this
function will allow us to analize the conformal properties of the model
in the domain $[g_N^{2 *},g_N^{2 **}]$.

In order to apply expression (\ref{4.9}) to the construction of the
$c(\tilde{g})$-function for the CGN model,
we shall begin by evaluating the generating functional for connected Green
Functions:
\beq
exp(-S_{eff}[\gamma])=\int Dg Dh exp(-S[\gamma]), \label{4.1}
\eeq
where $S[\gamma]$ is the action in the r.h.s of (\ref{2.16})
in an arbitrary euclidean metric:

\beqn
\lefteqn{S[\gamma]=kW[g]+kW[h^{-1}]}  \nonumber \\
& & +\frac{k}{4\pi}\tilde{g}\int d^2x\sqrt{\gamma}
(\gamma^{\mu\nu}+i\epsilon^{\mu\nu})
tr(g^{-1}\partial_{\mu}gh^{-1}\partial_{\nu}h),\label{4.2}
\eeqn
and $\gamma^{\mu\nu}$ is a background metric.
Then, differentiation of the effective action
with respect to  $\gamma^{\mu\nu}$ will give all connected
correlation functions containing products of
energy momentum tensor operators.
This will allow us to apply eq(3.1) to obtain $c(\tilde{g})$.

For the two point functions of energy momentum tensor operators
we are interested in, we have, up to contact terms (seagulls):

\beqn
\lefteqn{B_{\mu\nu\rho\sigma}[\gamma;x,y)=-\frac{2}{\sqrt{\gamma(x)}}
\frac{2}{\sqrt{\gamma(y)}}\frac{\delta^{(2)}
S_{eff}[\gamma]}{\delta \gamma^{\mu\nu}(x)\delta \gamma^{\rho\sigma}(y)}
|_{\gamma^{\mu\nu}=\delta^{\mu\nu}}=} \nonumber \\
& & <T_{\mu\nu}(x)T_{\rho\sigma}(y)>-<T_{\mu\nu}(x)><T_{\rho\sigma}(y)>
\nonumber \\
& & -\frac{2}{\sqrt{\gamma(x)}}\frac{2}{\sqrt{\gamma(y)}}<\frac{\delta^{(2)}
S[\gamma]}{\delta \gamma^{\mu\nu}(x)\delta
\gamma^{\rho\sigma}(y)}>|_{\gamma^{\mu\nu}=\delta^{\mu\nu}}.
\label{4.3a}
\eeqn

In order to evaluate  $S_{eff}$ as given by eq.(3.3),
we follow Ref.\cite{14}, and use the background field method.
We start by writing:
\beqn
g(x)&=&g_0~e^{i\pi(x)},  \nonumber \\
h(x)&=&h_0~e^{i\hat{\pi}(x)},  \label{4.3b}
\eeqn
with constant background fields $g_0$ and $h_0$.
Up to the one loop level we have to retain terms quadratic
in the fluctuation
fields $\pi$ and $\hat{\pi}$, so that the one loop effective action is given
by:

\beqn
\lefteqn{exp(-S_{eff}^{(1)}[\gamma])=
\int D\pi D\hat\pi} \nonumber \\
& & exp\left(\frac{k}{8\pi}\int d^2x
\sqrt{\gamma} \gamma^{\mu\nu}
tr(\partial_{\mu}\pi\partial_{\nu}\pi+
\partial_{\mu}\hat\pi\partial_{\nu}\hat\pi+
2\tilde{g} \partial_{\mu}\pi\partial_{\nu}\hat\pi)\right),
\label{4.4}
\eeqn
where $D\pi D\hat\pi$ is the product of the usual translationally invariant
measures (and $k=-(2N+1)$).

The path-integral (3.7) can be easily evaluated by changing variables from
$\pi$, $\hat\pi$ to:

\beqn
\pi'&=&\frac{(\pi+\hat\pi)}{\sqrt{2}}, \nonumber \\
\hat\pi'&=&\frac{(-\pi+\hat\pi)}{\sqrt{2}}. \label{4.5}
\eeqn
The resulting propagators are:

\beqn
<{\pi'}^a(x){\pi'}^b(y)>&=&-\frac{2\pi}{k} \frac{1}{1+
\tilde{g}}\delta^{ab} G(x,y)
\nonumber \\
<{\hat{\pi}'}^a(x){\hat{\pi}'}^b(y)>&=&-\frac{2\pi}{k} \frac{1}{1-\tilde{g}}
\delta^{ab} G(x,y)
\nonumber \\
<{\pi'}^a(x){\hat{\pi}'}^b(y)>&=&0 \label{4.6}
\eeqn
where $G(x,y)$ is the Green function for the Laplacian:
\beq
\partial_{\mu}(\sqrt{\gamma}\gamma^{\mu\nu}\partial_{\nu})G(x,y)=\delta(x-y).
\label{4.6a}
\eeq
As we stated before, $\tilde{g}$ will play the role of an effective coupling
constant moving (for fixed $\alpha$ i.e. for a given regularization
prescription) between $\tilde{g}=-1$ and $\tilde{g}=0$.
This interval corresponds to the variation
of the running coupling constant of the CGN Model $g^2_N$ between $g^{2
*}_N$ and $g^{2 **}_N$ (see eqs.(2.21) and (2.23)).

We then have:
\beqn
\lefteqn{exp(-S_{eff}^{(1)}[\gamma])=} \nonumber \\
& & \int D\pi' D\hat\pi' exp(-\frac{k}{4\pi}\int
d^2x \sqrt{\gamma}tr((1+\tilde{g})\pi' \nabla \pi'+(1-\tilde{g})\hat\pi'
\nabla \hat\pi')= \nonumber \\
& & [det(1+\tilde{g})\nabla det(1-\tilde{g})\nabla]^{-\frac{(N^2-1)}{2}}.
\label{4.7}
\eeqn
so that finally:
\beq
S_{eff}^{(1)}=\frac{N^2-1}{2}ln~det(1+\tilde{g})\nabla+
\frac{N^2-1}{2}ln~det(1-\tilde{g})\nabla. \label{4.8a}
\eeq

It must be stressed that the singularities in the
$\pi'$-propagator (eq.(3.9)) and in the first term in the effective
action in (\ref{4.8a}) when $\tilde{g}=-1$, were already present
in the original
model. In fact, for this value of the coupling constant, the WZW-sector  can
be written as in (\ref{2.19a})
and in this case we have an infinite factor that must be regularized.
We shall bypass this problem in what follows by excluding $\tilde{g}=-1$ and
consider this case at the end of our discussion.

With this in mind  and  noting that the constants $1+\tilde{g}$ and
$1-\tilde{g}$ do not affect the metric dependence of the effective
action (\ref{4.8a}), we have up to one loop:
\beq
S_{eff}^{(1)}= 2(N^2-1)D\{\gamma\}
,~~~if~-1<\tilde{g}\leq 0 .
\label{4.8b}
\eeq
with $D\{\gamma\}\equiv \frac{1}{2}ln~det\nabla$, given by\cite{15}:

\beqn
\lefteqn{D\{\gamma\}= -\frac{1}{48\pi}}\nonumber \\
& & \int d^2x d^2y \sqrt{\gamma(x)}
\sqrt{\gamma(y)}R(x)R(y)G(x,y)+ c\int d^2x \sqrt{\gamma(x)}. \label{4.8c}
\eeqn

We are now ready to evaluate the  one loop contribution to the c-function.
{}From the metric dependence of the laplacian operator determinant
(\ref{4.8c}) we obtain, in the plane-metric limit
$\gamma^{\mu\nu}=\delta^{\mu\nu}$:
\beqn
\lefteqn{ B_{\mu\nu\rho\sigma}[\gamma;x,y)=2(N^2-1)
\left( \frac{1}{24\pi}\delta_{\mu\nu}
(\partial_{\rho}\partial_
{\sigma}\delta(x-y)-
\delta_{\rho\sigma}\nabla_y\delta(x-y))-\right. } \nonumber \\
& & ~~~~~~~~~~~~~~\left. \frac{1}{24\pi}
(\partial_{\rho}\partial_{\sigma}\partial_{\mu}\partial_{\nu}(\frac{ln|x-y|}
{2\pi})-\delta_{\rho\sigma}\partial_{\mu}\partial_{\nu}\delta(x-y))\right),
\label{4.11}
\eeqn
for $-1<\tilde{g}\leq 0$.

Moreover, in the plane-metric limit, the v.e.v of $T_{\mu\nu}$ is
zero\cite{15},
and then (up to contact terms) we have:
\beqn
<T(x)T(0)>&=&\frac{1/2}{z^4}
2(N^2-1),~~if~-1<\tilde{g}\leq 0.
\nonumber \\
<T(x)\Theta(0)>&=&0,     \nonumber \\
<\Theta(x)\Theta(0)>&=&0. \label{4.12}
\eeqn

This leads to:
\beq
c(\tilde{g})^{(1)}=
2(N^2-1),~~if~-1<\tilde{g}\leq 0 .
\eeq
We then see that
at the one-loop level the $c$-function remains constant in all the interval
$[\tilde{g}^{*},\tilde{g}^{**}]$. Thus, up to this order, the effective action
behaves as if the theory was conformally invariant irrespectively of
the value of $\tilde{g}$, since the corresponding one-loop c-function
is scale independent. (We have also verified up to this order the vanishing
of the $\beta$-function).
We shall see however that at the two-loop level the c-function has a
non-trivial
scale dependence.

In order to compute $c(\tilde{g})$ up two-loop order, we must take into account
those vertices containing
three and four fields $\pi$ and $\hat\pi$. In terms of
the new variables (eq.(3.8)), they are given respectively by:
\beqn
\lefteqn{S^{(3)}=-\frac{ik}{12\sqrt2 \pi}\int
d^2x\epsilon^{\mu\nu}f^{abc}}\nonumber \\
& & \{(6\tilde{g}-2)
\partial_{\mu}\hat\pi'^a\partial_{\nu}\hat\pi'^b\hat\pi'^c-
6(\tilde{g}+1)\partial_{\mu}\pi'^a\partial_{\nu}\pi'^b\hat\pi'^c\}
\label{4.13}
\eeqn
and

\beqn
S^{(4)}= \frac{k}{24\pi}\int d^2x\sqrt{\gamma}\gamma^{\mu\nu}f^{abg}f^{cdg}
\{(1+\tilde{g})\partial_{\mu}\pi'^a\pi'^b\partial_{\nu}\pi'^c\pi'^d+
\nonumber \\
(1-7\tilde{g})\partial_{\mu}\hat\pi'^a\hat\pi'^b\partial_{\nu}\hat\pi'^c\hat
\pi'^d+
2(1-3\tilde{g})\partial_{\mu}\pi'^a\pi'^b\partial_{\nu}\hat\pi'^c\hat\pi'^d+
\nonumber \\
2(1+3\tilde{g})\partial_{\mu}\hat\pi'^a\pi'^b\partial_{\nu}\pi'^c\hat\pi'^d+
(1+7\tilde{g})\partial_{\mu}\pi'^a\hat\pi'^b\partial_{\nu}\pi'^c\hat\pi'^d+
\nonumber \\
(1-\tilde{g})\partial_{\mu}\hat\pi'^a\pi'^b\partial_{\nu}\hat\pi'^c\pi'^d\}
\label{4.14}
\eeqn

The two loop contribution to the effective action is then given by:

\beqn
\lefteqn{S_{eff}^{(2)}=<S^{(4)}>-\frac{1}{2}<S^{(3)}S^{(3)}>=}
\nonumber \\
& & \frac{\pi}{6k}N(N^2-1)\left([\frac{2}{1+\tilde{g}}+
\frac{(1-7\tilde{g})}{(1-\tilde{g})^2}+
\frac{(1+7\tilde{g})}{(1-\tilde{g}^2)}]\right.\times \nonumber \\
& & \int d^2x\sqrt{\gamma}\gamma^{\mu\nu}
(\partial_{\mu}^x
\partial_{\nu}^yG(x,y)G(x,y)-\partial_{\mu}^xG(x,y)\partial_{\nu}^yG(x,y))
|_{x=y}-\nonumber \\
& &  [\frac{(6\tilde{g}-2)^2}{2(1-\tilde{g})^3}+
\frac{6}{(1-\tilde{g})}]\times \nonumber \\
& & \left. \int d^2xd^2y
\epsilon^{\mu\nu}\epsilon^{\rho\sigma}\partial_{\mu}^x\partial_{\rho}^yG(x,y)
\partial_{\nu}^x\partial_{\sigma}^yG(x,y)G(x,y)\right)
\label{4.15}
\eeqn

This expression is ill-defined since it has infrared (IR) and ultraviolet
(UV) divergences. To tame
IR divergences, one can add a mass term as in Ref.\cite{10}, but
this change does not affect the renormalization group properties
of the model. (We briefly discuss how one handles these divergences in the
Appendix).

In order to eliminate UV-divergences, we use the
dimensional regularization method in the presence of a background metric
\cite{lu}. We discuss this issue in the Appendix and just
quote here the final form of the regularized two-loop contribution to the
effective action. It is given by:

\beq
S_{eff}^{(2)}|_{REG}=f(\tilde{g})\frac{1}{k}N(N^2-1)
\left[ \frac{\pi}{6}\int d^2x \sqrt{\gamma}
(\gamma^{\mu\nu}\partial_{\mu}\overline{G}
\partial_{\nu}\overline{G}-\frac{R}{2\pi}\overline{G})\right]
\label{4.17}
\eeq
with $f(\tilde{g})$ given by:
\beq
f(\tilde{g})=\frac{6\tilde{g}^3-6\tilde{g}^2-4\tilde{g}+2}
{(1-\tilde{g})^3(1+\tilde{g})},
\label{4.18}
\eeq
and $\overline{G}(x)$ is the finite part of $G(x,y)$ at $x=y$ (\ref{A.4}):

\beq
\overline{G}(x)\equiv \left[G(x,y)+\frac{1}{2\pi}ln~s(x,y)\right]|_{x=y}
\label{4.18a}
\eeq

Using conformal coordinates, it can be shown that the term in brackets
in (\ref{4.17}) coincides with $D\{\gamma\}$ as given by eq.(\ref{4.8c}):
\beq
D\{\gamma\}=\frac{1}{2}ln~det\nabla=
\frac{\pi}{6}\int d^2x \sqrt{\gamma}(\gamma^{\mu\nu}\partial_{\mu}\overline{G}
\partial_{\nu}\overline{G}-\frac{R}{2\pi}\overline{G}),
\label{4.19}
\eeq
and then  the two-loop contribution to the effective action only
amounts to a change in the multiplicative factor
appearing in the r.h.s of (\ref{4.8b}).

Then, using (\ref{4.11}), we have that the two loop contribution
to the c-function is given by:

\beq
c(\tilde{g})=(N^2-1)
\left( 2-\frac{N}{k}f(\tilde{g})\right) ~~~if~-1<\tilde{g}\leq 0.
\label{4.22}
\eeq

In the fixed point $\tilde{g}=0$ we have
\beq
c(0)=2(N^2-1)(1-\frac{N}{k}).
\label{4.22'}
\eeq
Note that at the critical point the VCC for two SU(N)$_k$-WZW Models, as
given by eq(\ref{2.21b}), has a $\frac{1}{k}$-expansion which coincides with
(\ref{4.22'}). This result is to be expected since $\frac{1}{k}$-expansion
at the critical point should coincide with loopwise expansion.
This result is in agreement with that obtained in Ref.\cite{14}, where a
perturbative evaluation of the central charge of a SU(N)$_k$-WZW Model in
the IR-stable fixed point was performed.

Also, $c(\tilde{g})$ given by (\ref{4.22}) is stationary at this point
reflecting the fact that it corresponds a to critical point for the theory
(i.e. the $\beta$-function vanishes at this point).

We now analize the case $\tilde{g}=-1$.
At this point the partition function can be written as in eq.(2.22), and we
have an overall infinite factor that must be carefully handled in the
perturbative evaluation of the effective action.

In our treatment this divergence manifests itself first as a simple pole
in the propagator for the $\pi'$-field (eq.(3.9)). The one-loop effective
action $S_{eff}^{(1)}$ given in eq.(3.12) reflects this fact in the singularity
of its first term which is divergent at $\tilde{g}=-1$. (Out of this point
$S_{eff}^{(1)}$ is $\tilde{g}$-independent and corresponds to the effective
action for $2(N^2-1)$ free bosons).

At the two-loop level we have again a result that is singular at
$\tilde{g}=-1$. In fact, the two-loop contribution to the effective action
$S_{eff}^{(2)}$ given in eq.(3.21) has a simple pole at this point.
We shall then define the effective action by choosing a regularization
giving the desired values of $c(g)$ also at the critical point
$\tilde{g}=-1$. (Of course it has to fullfill the condition of stationarity
at this point)

This amounts to define:

\beq
S_{eff}=(N^2-1)D\{\gamma\}\left\{
\begin{array}{lc}
1-\frac{N}{k}&if~\tilde{g}=-1\\
{}~~~~~&~~~~~~\\
{}~~~~~&~~~~~~\\
2-\frac{N}{k}f(\tilde{g})|_{Reg}&if~-1<\tilde{g}\leq 0,
\end{array} \right.
\eeq
where:
\beq
f(\tilde{g})|_{Reg}=\frac{-\tilde{g}^3+21\tilde{g}^2-12\tilde{g}+4}
{2(1-\tilde{g})^3}
\label{4.21}
\eeq

Finally the two-loop Zamolodchikov's c-function is given by:

\beq
c(\tilde{g})=(N^2-1)\left\{
\begin{array}{lc}
1-\frac{N}{k}&if~\tilde{g}=-1\\
{}~~~~~&~~~~~~\\
{}~~~~~&~~~~~~\\
2-\frac{N}{k}f(\tilde{g})|_{Reg}&if~-1<\tilde{g}\leq 0.
\end{array} \right.
\eeq

This function is monotonically decreasing with the running coupling constant
$g^2_N$ through (3.2),
and has no additional stationary points in the domain
$[\tilde{g}^{*},\tilde{g}^{**}]$
thus confirming that the model becomes
conformally~ invariant only~ for those values of the
coupling constant obtained
in Ref.\cite{2} (see eqs.(2.21),(2.23)).

Of course, this is valid in the framework of the adopted
regularization scheme.
Regularization scheme dependence of the Zamolodchikov's c-function has
been discussed in the context of a critical theory perturbed by a relevant
operator in Ref.\cite{Boya}. In particular
it was shown in that paper that conditions
a), b) and c) listed in the begining of the present section are scheme
independent while the relation between the
$\beta$-function and the gradient of the $c$-function depends on the
regularization scheme.

Concerning conditions a), b) and c) they are in fact fullfilled by our
expression for the $c$-function
(eq.(3.29)) in the region $-1<\tilde{g}\leq 0$, but we have a finite
discontinuity in $\tilde{g}=-1$.
This discontinuity has its origin in the infinite overall factor in the
expression for the partition function of the model (see eq.(2.22)).
This factor can be eliminated by quotienting eq.(2.22) by the integral
over the group.
Another interesting issue concerning the study of the c-function
is the analysis of Frishman et al
in Ref.\cite{yan} where general arguments were given for the case in which
there exist
two critical points, to show that there would not be mass generation between
these points. We have shown the non existence of additional stationary points
between
those encountered in Ref.\cite{2} this showing that the conformal symmetry
remains broken between these points even if there is no mass generation
as stated in Ref.\cite{yan}.

As we mentioned in the introduction,
there exists previous calculations of the $c$-function like in the (exactly
solvable) Schwinger Model \cite{chinos}.
Our calculation provides the first perturbative test in a field theoretical
model of the Zamolodchikov's approach to the conformal behaviour of two
dimensional systems. Moreover, it has the additional interest of probing
a renormalization group flow between two {\it non-trivial} critical points.

To summarize, we have computed Zamolodchikov's c-function up to two loops
for the Chiral Gross-Neveu Model. We have shown that:

a) It gives the desired value for the central charge in both critical points
obtained in Ref.\cite{2}.

b) It is stationary at these points.

c) It is monotonically increasing with the running coupling constant
$\tilde{g}$ in the domain $-1<\tilde{g}\leq 0$.

d) It shows no other critical points in this domain.


The evaluation of the Zamolodchikov's c-function is
an important step towards the study
of a theory away from criticality but it could
be also useful from the knowledge of the values of the Virasoro central
charge  at the critical point, (a non-perturbative result),
to calibrate approximation schemes to be applied
in the evaluation of other correlation functions in the model.

The approach used in this paper can be also applied to study minimal
coset theories in the FQS series. It was shown in Ref.\cite{M} that starting
from a critical theory in this series and perturbing
the system adequately, the model flows to the following theory in the series.
It would be then interesting to study the flow between adyacent minimal
models in the coset formulation given in Refs.\cite{Kara} since one has to
study
WZW theories with a perturbation that can be constructed from the original
fields. Work on these aspects is in progress.
\vspace{2cm}

\noindent {\bf ACKNOWLEDGEMENTS}

I am grateful to F.A.Schaposnik for suggesting the problem
and for useful discussions and
encouragement at the various stages of the realization of this work. I am
also indebted to G.L.Rossini for valuable comments.

\appendix

\section{Dimensional regularization of the two-loop effective
action}
\setcounter{equation}{0}

In this appendix we describe how divergences are regularized in the two-loop
contribution to the effective action (\ref{4.15}). Concerning infrared
divergences (IR),
they can be eliminated, for example, adding a term of the form\cite{10}:

\beq
S_m(g,h)=\frac{-m^2k}{16\pi}\int
d^2x\left(tr(g+{g}^{\dagger})+tr(h+{h}^{\dagger})\right),
\label{A.0}
\eeq
which, when $g$ and $h$ are expressed as in (\ref{4.3b}), will supply the mass
through the terms quadratic in $\pi$ and $\hat\pi$ .
Now, these terms do not affect renormalization group coefficients such
as $\beta$ or $c(g)$, and
it can be shown that observables invariant under the action of the group will
have IR perturbative expansions\cite{?}.
In fact it can be shown explicitly seen that the presence of a mass term does
not change the short distance behaviour of the effective action and then
its metric dependence is not affected.
Let us now discuss UV divergences.
The two-loop effective action is given by:
\beqn
\lefteqn{S_{eff}^{(2)}=<S^{(4)}>-\frac{1}{2}<S^{(3)}S^{(3)}>=}
\nonumber \\
& & \frac{\pi}{6k}N(N^2-1)\left([\frac{2}{1+\tilde{g}}+
\frac{(1-7\tilde{g})}{(1-\tilde{g})^2}+
\frac{(1+7\tilde{g})}{(1-\tilde{g}^2)}]\times \right. \nonumber \\
& & \int d^2x\sqrt{\gamma}\gamma^{\mu\nu}
(\partial_{\mu}^x
\partial_{\nu}^yG(x,y)G(x,y)-\partial_{\mu}^xG(x,y)\partial_{\nu}^yG(x,y))
|_{x=y}-\nonumber \\
& &  [\frac{(6\tilde{g}-2)^2}{2(1-\tilde{g})^3}+
\frac{6}{(1-\tilde{g})}] \times \nonumber \\
& & \left.\int d^2xd^2y
\epsilon^{\mu\nu}\epsilon^{\rho\sigma}\partial_{\mu}^x\partial_{\rho}^yG(x,y)
\partial_{\nu}^x\partial_{\sigma}^yG(x,y)G(x,y)\right).
\label{A.1}
\eeqn

In order to regularize the divergent integrals in the r.h.s. of (\ref{A.1}),
we use the dimensional regularization method, adapted so as to take into
account the metric dependence of the effective action\cite{lu}-\cite{14}.

Consider in (A.2):
\beq
{\cal I}_1=\int d^2x\sqrt{\gamma}\gamma^{\mu\nu}
(\partial_{\mu}^x
\partial_{\nu}^yG(x,y)G(x,y)-\partial_{\mu}^xG(x,y)\partial_{\nu}^yG(x,y))
|_{x=y}.
\label{A.2}
\eeq

We have for the regularized propagator at coinciding arguments:

\beq
G(x,x)=\overline{G}(x)+\frac{1}{2\pi\epsilon},
\label{A.3}
\eeq
where $\epsilon=d-2$ and $\overline{G}(x)$ is the finite part of
the propagator defined as

\beqn
\overline{G}(x,y)=G(x,y)+\frac{1}{2\pi}ln~s(x,y) \nonumber \\
\overline{G}(x)=\overline{G}(x,y)|_{x=y}. \label{A.4}
\eeqn
($ln~s(x,y)$ is the geodesic distance between the two points, and control
the short distance singularity of the Green function $G(x,y)$).

For the first derivative, using $G(x,y)=G(y,x)$, one can show
that:

\beq
\partial_{\mu}^xG(x,y)|_{x=y}=\partial_{\mu}^yG(x,y)|_{x=y}=\frac{1}{2}
\partial_{\mu}\overline{G}(x),
\label{A.5}
\eeq
For the second derivative one has:
\beq
\partial_{\mu}^x\partial^{y\mu}G(x,y)|_{x=y}=-\frac{R}{8\pi}.
\label{A.6}
\eeq
where $R(x)$is the scalar curvature.

Hence, the regularized expression is:

\beq
{\cal I}_1|_{Reg}=-\int d^2x\sqrt{\gamma(x)} \left(\frac{1}{4}
\partial_{\mu}\overline
{G}(x)\partial^{\mu}\overline{G}(x)+\frac{1}{8\pi}R(x)\overline{G}(x)\right),
\label{A.7}
\eeq
This expression can be written in terms of the regularized expression for the
determinant of the laplacian operator given in eq.(\ref{4.8c}):
\beq
{\cal I}_1|_{Reg}=\frac{9}{2\pi}D\{\gamma\}.
\label{A.8}
\eeq

In order to regularize the second divergent integral in the r.h.s of
eq.(\ref{A.1}) given by:

\beq
{\cal I}_2=\int d^2xd^2y
\epsilon^{\mu\nu}\epsilon^{\rho\sigma}\partial_{\mu}^x\partial_{\rho}^yG(x,y)
\partial_{\nu}^x\partial_{\sigma}^yG(x,y)G(x,y),
\label{A.9}
\eeq
we shall make use of conformal coordinates. In this coordinates, the metric
takes the form:
\beq
\gamma_{\mu\nu}(x)=e^{w(x)}\delta_{\mu\nu},
\label{A.10}
\eeq
and ${\cal I}_2$ becomes:

\beq
{\cal I}_2=\frac{1}{(2\pi)^3}\int d^2xd^2y \epsilon^{\mu\nu}
\epsilon^{\rho\sigma}\partial_{\mu}^x\partial_{\rho}^yln|x-y|~
\partial_{\nu}^x\partial_{\sigma}^yln|x-y|~ln|x-y|.
\label{A.11}
\eeq

We eliminate the product of totally antisymmetric tensors
$\epsilon^{\mu\nu}$ via
\beq
\epsilon^{\mu\nu}\epsilon^{\rho\sigma}=\gamma^{\mu\sigma}\gamma^{\nu\rho}
-\gamma^{\mu\rho}\gamma^{\nu\sigma},
\label{A.12}
\eeq
Then, after integration by parts ${\cal I}_2$ can be written as:

\beq
{\cal I}_2=-\frac{1}{2}\int d^2x \left(\partial_{\mu}^xG(x,y)
\partial^{y\mu}G(x,y)\right)|_{x=y}.
\label{A.13}
\eeq

Finally using eq.(\ref{A.5}) we get:

\beq
{\cal I}_2|_{Reg}=\frac{3}{4\pi}D\{\gamma\}.
\label{A.14}
\eeq

The whole two-loop contribution to the renormalized
effective action $S_{eff}^{(2)}$ is then given by:

\beq
S_{eff}^{(2)}|_{REG}=g(\tilde{g})\frac{1}{k}N(N^2-1)D\{\gamma\},
\label{A.15}
\eeq
with:
\beq
g(\tilde{g})=\frac{6\tilde{g}^3-6\tilde{g}^2-4\tilde{g}+2}
{(1-\tilde{g})^3(1+\tilde{g})}.
\label{A.16}
\eeq

\end{document}